\documentclass[prl,nofootinbib,twocolumn,showpacs,preprintnumbers]{revtex4}
\usepackage[a4paper, hdivide={1.91cm,,1.165cm}, vdivide={1.83cm,,3.4cm}]{geometry}

\usepackage{amstext,amssymb}
\usepackage[intlimits]{amsmath}
\usepackage[english]{babel}
\usepackage[ansinew]{inputenc}
\usepackage{graphicx}
\usepackage{units}
\usepackage{hyperref}

\hypersetup{
  pdftitle={How Stable is the Photon?},
  pdfauthor={Julian Heeck}
}

\begin{document}

\newcommand{\del}{\partial}
\newcommand{\dd}{\mathrm{d}}

\title{How Stable is the Photon?}

 \author{Julian \surname{Heeck}}
 \email{julian.heeck@mpi-hd.mpg.de}
 \affiliation{Max-Planck-Institut f\"ur Kernphysik, Saupfercheckweg 1, 69117 Heidelberg, Germany}


\preprint{Phys.\ Rev.\ Lett.\  {\bf 111}, 021801 (2013)}

\begin{abstract}

Yes, the photon. While a nonzero photon mass has been under experimental and theoretical study for years, the possible implication of a finite photon lifetime lacks discussion. The tight experimental upper bound of the photon mass restricts the kinematically allowed final states of photon decay to the lightest neutrino and/or particles beyond the Standard Model. We discuss the modifications of the well-measured cosmic microwave background spectrum of free streaming photons due to photon mass and lifetime and obtain model-independent constraints on both parameters---most importantly a lower direct bound of $\unit[3]{yr}$ on the photon lifetime, should the photon mass be at its conservative upper limit. In that case, the lifetime of microwave photons will be time-dilated by a factor order~$10^{15}$.

\end{abstract}

\maketitle


Classical electrodynamics, as encoded in Maxwell's equations, can be readily extended to allow for a nonzero photon mass; the resulting Proca equations~\cite{proca} then describe the behavior of a massive spin-1 field, and have since been used to set an impressive upper limit on the photon mass of $m < \unit[2\times 10^{-54}]{kg}$~\cite{photonmass}, or $ \unit[10^{-18}]{eV}$ in the natural units used in this Letter ($\hbar = c = k_B = 1$). A nonzero photon mass is often dismissed on theoretical grounds, as the insertion of a mass term to the Lagrangian of quantum electrodynamics (QED) breaks gauge invariance and might therefore spoil renormalizability, i.e.~the consistency of the theory at quantum level. This is, however, not the case as the Proca Lagrangian can be viewed as a gauge-fixed version of the St\"uckelberg Lagrangian~\cite{Stueckelberg:1900zz}, which restores gauge invariance. For an exhaustive review we refer to~\cite{Ruegg:2003ps}. To the point: gauge bosons of abelian symmetries are permitted a 
mass by means of the St\"uckelberg mechanism---retaining gauge invariance, unitarity, and renormalizability.

The question of a photon mass in QED is then purely experimental, as there is no theoretical prejudice against a small $m$ over $m=0$.\footnote{A small $m$ is technically natural~\cite{tHooft:1979bh}, in that all radiative corrections are proportional to $m$.} However, we already know that QED is just the low-energy approximation of the Glashow--Weinberg--Salam model of electroweak interactions, so our above motivation for the photon mass might be in danger. Fortunately, the electroweak gauge group $SU(2)_L\times U(1)_Y$ still features an abelian factor---the hypercharge $U(1)_Y$---that can be used in a St\"uckelberg mechanism. The resulting mass for the hypercharge gauge boson eventually generates again a massive photon~\cite{Kuzmin:2001pg}.\footnote{The same trick works, for example, in simple left--right symmetric models~\cite{left-right}, where the hypercharge $U(1)_Y$ itself results from the breakdown of $SU(2)_R\times U(1)_{B-L}$: A St\"uckelberg mass of the $B-L$ boson trickles down and makes the 
photon massive.} A detailed discussion of this procedure and its implications can be found in Ref.~\cite{Ruegg:2003ps}. Since the St\"uckelberg mechanism only works for abelian groups, the grand unification of the Standard Model (SM) gauge group $SU(3)_C\times SU(2)_L \times U(1)_Y$ into a simple nonabelian group like $SU(5)$, $SO(10)$, or $E_6$ would necessarily result in a truly massless photon~\cite{McKeon:2006ym}. Turning this around, the discovery of a massive photon would exclude a huge number of grand unified theories---and, obviously, be a spectacular finding in its own right.

Let us now move on to the key point of this Letter: If one can constrain the mass of a photon, one should also be able to constrain its lifetime.
Massless photons in QED are stable purely due to kinematical reasons, there are no additional quantum numbers that forbid a decay. Recalling the tight upper bound on the photon mass though, there are not many possible final states---indeed, only one known particle could be even lighter than the photon: the lightest neutrino $\nu_1$. This is because current neutrino-oscillation experiments can only fix the two mass-squared differences $\Delta m_{31}^2 = m_3^2-m_1^2$ and $\Delta m_{21}^2 = m_2^2-m_1^2$ of the three neutrinos, so the absolute mass scale is not known as of yet~\cite{Luo:2012fm}.
Kinematically, this opens up the possibility of a decay $\gamma \rightarrow \nu_1\nu_1$---should $m_1 < m/2$ hold.\footnote{The naive prototype model---augmenting the SM by only two right-handed neutrinos (SM$+2\nu_R$)---is problematic, as the initially massless $\nu_1$ will unavoidably pick up a finite mass at loop level~\cite{Petcov:1984nz}, which can be too large for our purposes~\cite{Davidson:2006tg}. Fine-tuned solutions aside, we can obtain a simple valid model by imposing a $B-L$ symmetry on the SM$+2\nu_R$, resulting in two Dirac neutrinos and one exactly massless Weyl neutrino.} 
This loop-suppressed process can be calculated in the SM (using e.g.~a seesaw mechanism to make neutrinos massive in a renormalizable way), and is of course ridiculously small~\cite{Hare:1972bq}---being suppressed by the small photon mass, the heavy particles in the loop and maybe the smallest neutrino mass, depending on the operator that induces this decay. We also note that one of the side effects of a massive hypercharge boson---besides a massive photon---are tiny electric charge shifts of the known (chiral) elementary particles~\cite{Kuzmin:2001pg,Ruegg:2003ps}. The neutrino then picks up a charge $Q_\nu \propto e \, m^2/M_W^2$, which gives rise to a correspondingly small tree-level decay rate $\gamma \rightarrow \nu_1\nu_1$.
Still, unmeasurable small SM rates never stopped anyone from looking for a signal, as it would be a perfect sign for new physics.

Particles beyond the SM could not only increase the rate $\gamma \rightarrow \nu_1\nu_1$, but also serve as final states themselves, as some extensions of the SM feature additional (close to) massless states; examples include sterile neutrinos, hidden photons, Goldstone bosons and axions. These weakly interacting sub-eV particles~\cite{Jaeckel:2010ni} are less constrained than neutrinos, and photon decay might be an indirect effect of these states. Although mainly of academic interest, we also mention that a massive photon provides the possibility of faster-than-light particles---and a decaying photon even predicts them. The question of photon decay is therefore obviously relevant even if the lightest neutrino turns out to be an inaccessibly heavy final state.

Following the above motivation, we set out to find limits on the photon mass $m$ and lifetime $\tau_\gamma$ as model-independent parameters. Most importantly, we do not care about the daughter particles for now. Because of the small allowed values for $m$, all measurable photons around us are highly relativistic, making a decay hard to observe because of time dilation. Correspondingly, a good limit on $\tau_\gamma$ needs a large number of low-energy photons from well-known far-away sources. Seeing as we have access to very accurate measurements of the cosmic microwave background (CMB)---consisting of the oldest photons in the visible Universe---we will take $m$ and $\tau_\gamma$ as parameters that will modify the blackbody radiation law---given by the Planck spectrum---and fit the CMB spectrum to obtain bounds on both parameters. Similar analyses have been performed to obtain a limit on the neutrino lifetime in the channels $\nu_i \rightarrow \gamma\, \nu_j$~\cite{Mirizzi:2007jd,Broggini:2012df}. In our case,
 we are, however, not looking for a spectral line on top of the CMB, but rather a diminished overall intensity and change of shape.

Before delving into the details, let us present a back-of-the-envelope estimate: CMB photons with low energies around meV have a lifetime $\tau = \gamma_\mathrm{L} \tau_\gamma$ that is increased by a relativistic Lorentz factor $\gamma_\mathrm{L} = E/m \simeq \unit[1]{meV}/\unit[10^{-18}]{eV} = 10^{15}$. This lifetime has to be compared to the age of the Universe $t_0 \simeq \unit[13.8\times 10^{9}]{yr}$ (or the corresponding comoving distance). Seeing as an improved accuracy $A$ in the measurements will increase the bound, we can estimate $\tau_\gamma \gtrsim t_0/\gamma_\mathrm{L}  A$. We therefore expect a lifetime constraint in the ballpark of years from the very precise CMB measurements ($A\simeq 10^{-4}$), which will be confirmed by the more refined analysis below.

The photon mass changes the spectral energy density of blackbody radiation to
\begin{align}
 \rho (T,E) \,\dd E= \frac{1}{\pi^2} \, \frac{E^3 \,\dd E}{e^{E/T}-1}\, \sqrt{1-\frac{m^2}{E^2}} \,,
\end{align}
because of the modified dispersion relation $p^2 = E^2-m^2$, but it is unclear how to include the decay width. The expansion of the Universe also needs to be taken into account, as the blackbody spectrum no longer stays in shape for $m\neq 0$. Let us therefore give a brief derivation of the energy spectrum of massive unstable photons during cosmic expansion.

Ignoring the width for a moment, the number density of massive photons right after decoupling (at the time of last scattering $t_L \simeq \unit[400\, 000]{years}$) is given by~\cite{Zhang:2008if}
\begin{align}
\begin{split}
 n_0 (p,t) \,\dd p &= \left( \frac{a(t_L)}{a(t)}\right)^3 n_0 (p_L, t_L) \, \dd p_L\\
&=  \frac{4 \pi g p^2 \, \dd p/(2\pi)^3}{\exp \left( \sqrt{ p^2 + m^2 \left( \frac{a(t_L)}{a(t)}\right)^2}/T\right) -1} \,,
\end{split}
\end{align}
where $p = p_L \ a(t_L)/a(t)$ is the redshifted momentum, $T$ the temperature at time $t$, and $g$ the number of spin states. We take $g=2$, because only the transverse modes are excited before decoupling (this implicitly constrains $m$, as discussed below). The chemical potential of massless photons is zero, and since we assume that as our initial condition at $t_L$, we set it to zero in all our calculations.

Including the width, we can write down the differential equation for the time evolution of the number density
\begin{align}
 \frac{\dd}{\dd t} n(p,t) =  \frac{\dd}{\dd t} n_0 (p,t) - \Gamma (p) n_0 (p,t) \,.
\end{align}
The first term on the right-hand side describes the number density dilution due to the expansion of the Universe, while the second one is due to photon decay. The width can be obtained from the rest-frame width $\Gamma_0 = 1/\tau_\gamma$ by a Lorentz boost: $\Gamma (p) \simeq \Gamma_0 \frac{m}{p}$. We use the boundary condition $n(p,t_L) = n_0 (p,t_L)$ and obtain the number density today
\begin{align}
 n(p, t_0) = n_0 (p,t_0)- \Gamma_0 \int_{t_L}^{t_0} \frac{m}{p} n_0 (p,t) \, \dd t \,.
\end{align}
The integral can be evaluated to
\begin{align}
\begin{split}
 \int_{t_L}^{t_0} \frac{m}{p} n_0 (p,t) \, \dd t &=  \frac{m}{p_L} n_0 (p_L ,t_L) \int_{t_L}^{t_0} \frac{a(t_L)}{a(t)}  \, \dd t\\
&= \frac{m}{p} n_0 (p ,t_0) \ d_L \,,
\end{split}
\end{align}
with the comoving distance of the surface of last scattering $d_L = \int_{t_L}^{t_0} a(t_0)/a(t)  \, \dd t \simeq 47$~billion lightyears.
Overall we have:
\begin{align}
\begin{split}
 n(p, t_0) &\simeq  n_0 (p,t_0)\left(1- \Gamma_0 \frac{m}{p}  d_L \right)\\
 &\simeq n_0 (p,t_0) \exp \left(- \Gamma_0 \frac{m}{p}  d_L \right)\,.
\end{split}
\end{align}
The energy density relevant for the CMB spectrum is then obtained by multiplying $n (p,t_0)$ with $E = \sqrt{p^2+m^2}$:
\begin{align}
\begin{split}
 \rho (E,T) \,\dd E &\simeq \frac{1}{\pi^2} \, \frac{E^3 \,\dd E}{e^{\sqrt{ E^2-m^2}/T}-1}\, \sqrt{1-\frac{m^2}{E^2}} \\
 &\quad \times 
\exp \left(- \Gamma_0 \frac{m}{E} d_L \right) \,,
\end{split}
\label{eq:spectrum}
\end{align}
where we approximated 
\begin{align}
 \sqrt{ p^2 + m^2 \left( \frac{a(t_L)}{a(t)}\right)^2} \simeq \sqrt{E^2-m^2}
\end{align}
because $a(t_L)/a(t_0) \simeq 8 \times 10^{-4}$.
Because of this approximation, the limit $\rho (E\rightarrow m,T)$ is nonzero, which is, however, of no importance for the CMB analysis in this Letter.

\begin{figure}[tb]
\setlength{\abovecaptionskip}{-2ex}
	\begin{center}
		\includegraphics[width=0.45\textwidth]{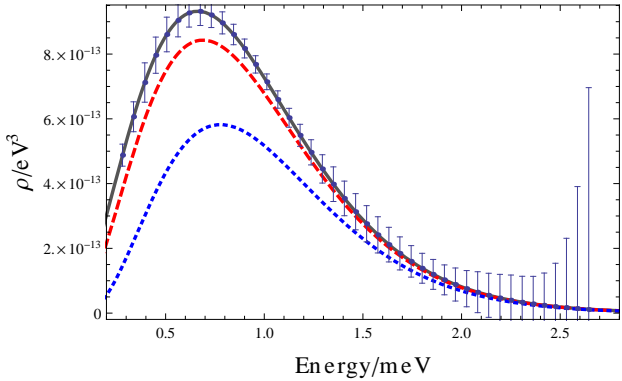}
	\end{center}
		\caption{CMB spectral distribution for $\Gamma_0 t_0 m = 0$ (gray), $\Gamma_0 t_0 m = \unit[2\times 10^{-5}]{eV}$ (dashed red line) and $\Gamma_0 t_0 m = \unit[10^{-4}]{eV}$ (dotted blue line) using Eq.~\eqref{eq:spectrum}, as well as the COBE data (error bars multiplied by $1000$ to be visible). In all cases the mass is $m<\unit[10^{-6}]{eV}$ and has no visible effect.}
	\label{fig:blackbody}
\end{figure}

Equation~\eqref{eq:spectrum} is the key equation of this Letter and will now be used to set constraints on $m$ and $\Gamma_0$. For illustrative purposes we show the spectrum for various values in Fig.~\ref{fig:blackbody}. As expected from time-dilation arguments, the low-energy part of the spectrum shows the strongest deviations, which fortunately also features the smallest error bars.

\begin{figure}[tb]
\setlength{\abovecaptionskip}{-2ex}
	\begin{center}
		\includegraphics[width=0.45\textwidth]{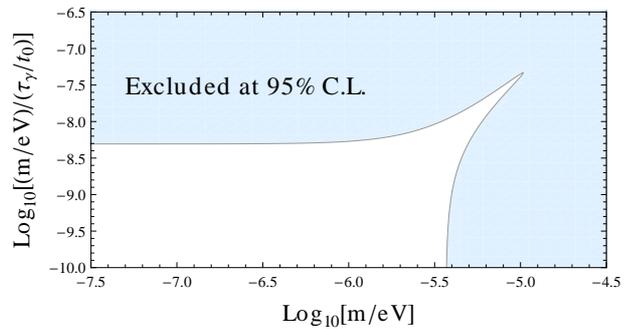}
	\end{center}
		\caption{Constraints on photon mass $m$ and lifetime $\tau_\gamma$ from the CMB spectrum.}
	\label{fig:cmbconstraint}
\end{figure}

Using the COBE (COsmic Background Explorer) data set of the CMB~\cite{Fixsen:1996nj} we can construct a simple $\chi^2$ function to fit the spectrum from Eq.~\eqref{eq:spectrum}.\footnote{Ground-based and balloon experiments probe the CMB down to energies $\sim\unit[10^{-6}]{eV}$, which typically have much larger errors. Additionally, there is an excess at low energies that is not understood yet~\cite{Fixsen:2009xn}, so we do not include those data.} The best fit values are at $m = 0 = \Gamma_0$, so we can only obtain exclusion ranges, shown in Fig.~\ref{fig:cmbconstraint}. The limit on the photon mass is not competitive with other experiments---$m < \unit[3\times 10^{-6}]{eV}$---but for the photon width we find the only existing (and model-independent) bound
\begin{align}
 \tau_\gamma >2\times 10^{-10} \left( \frac{m}{\unit[10^{-18}]{eV}}\right) \, t_0
\label{eq:lifetime}
\end{align}
at $95\%$~C.L. This would correspond to a photon lifetime of only three years, should the photon mass be close to its current bound. 
Another useful form of the constraint is given by
\begin{align}
 \left( \frac{m}{\unit[10^{-18}]{eV}}\right)\left( \frac{\Gamma_0}{\unit[7.5\times 10^{-24}]{eV}}\right) < 1\,.
 \label{eq:bound2}
\end{align}
For two-particle fermionic final states $X$, the decay rate $\gamma \rightarrow X X$ from (effective) interactions like $g \overline{X}\gamma_\mu X A^\mu$ will be of the form $\Gamma_0 \sim g^2 m/4\pi$~\cite{Hare:1972bq}. With Eq.~\eqref{eq:bound2} we can constrain $g\lesssim 0.03 \,e$, which corresponds to a very large effective electric charge and is excluded by other experiments~\cite{Davidson:2000hf}.\footnote{It is of course trivial to reinterpret bounds on millicharged particles~\cite{Davidson:2000hf} in terms of photon decay.} In particular, final state neutrinos are far better constrained by their electric properties (see, e.g., Ref.~\cite{Broggini:2012df} for a recent review) to be relevant in photon decay. 
Our complementary and model-independent approach should be interesting nonetheless, as it constitutes the only direct constraint on the photon lifetime as of yet.

Let us make a couple more comments to illustrate that our analysis above is somewhat inconsistent. Our approach basically assumed a vanishing or negligible number density of St\"uckelberg scalars $\phi$ and daughter particles $X$ prior to photon decoupling. To ensure this, $m$ and $\Gamma_0$ need to be small: $\phi$ has only the interaction $m A^\mu \del_\mu \phi$, so for small mass $m$, it will not be in equilibrium with the rest of the SM. The creation rate of $\phi$ via $e \gamma \leftrightarrow e \phi$ is proportional to $\alpha^2 m^2/T$, which has to be smaller than the expansion rate of the Universe $H(T) \simeq T^2/M_\mathrm{Pl}$---at least before weak decoupling around $T\simeq \unit[1]{MeV}$---in order to not put $\phi$ in thermal equilibrium at Big Bang nucleosynthesis (BBN). For $m< \unit[10^{-3}]{eV}$, only the transversal polarizations of the photon are excited, making it okay to treat the photon as massless before BBN. For the initial condition of our blackbody calculation however, we need to 
ensure that only the two transverse degrees of freedom of the photon are excited at the surface of last scattering at $T\simeq \unit[0.25]{eV}$. This requires $m< \unit[5\times 10^{-13}]{eV}$, making our approach a little inconsistent, because at these low masses the primordial plasma---consisting mainly of partly ionized hydrogen and helium---cannot be ignored. We will remark on this below.

On to the daughter particles: the interaction rate of photons with their will-be daughter particles at temperature $T$ will be something like $\Gamma_0 T/m$, as it should be finite in the limit $m\rightarrow 0$. This rate has to be smaller than the expansion rate of the Universe at BBN---unless the final daughter particles are neutrinos.
This gives the condition $\Gamma_0 \ll 10^{-22} m < \unit[10^{-40}]{eV}$, which is far stronger than the bound we obtained from the CMB analysis above, directly related to the fact that the minicharge of new ultralight particles is tightly constrained~\cite{Davidson:2000hf}. One should be careful with the above constraint though, because additional degrees of freedom at BBN are still allowed by cosmological observations~\cite{Ade:2013lta}.

Having discussed the initial conditions of our analysis---which degrees of freedom are present at recombination---it is time to scrutinize our main assumption: that the photons are free streaming. This is usually a very good approximation, as the density of ionized hydrogen is rather small after recombination, but it is still large enough to induce a plasma mass as large as $\unit[10^{-9}]{eV}$ to the photon. Further complications arise from the nonionized hydrogen and helium, as they effectively make the Universe a refractive medium---changing the dispersion relation of on-shell photons even further. This has been emphasized in Ref.~\cite{Mirizzi:2009iz}, where CMB constraints on photon oscillations into hidden photons~\cite{Georgi:1983sy} have been discussed. Their analysis (and phenomenology) is very similar to our discussion of photon decay, but in our case the inclusion of the plasma is more difficult. The photon in a medium requires a careful treatment, as it becomes just one of several quasiparticles 
that can be excited. (A well-studied example relevant to our discussion is the decay of plasmons---effectively massive photons---into neutrinos as a mechanism to cool stars~\cite{Adams:1963zzb,Broggini:2012df}.) This makes it difficult, if not impossible, to constrain the properties of a free photon---namely, $m$ and $\tau_\gamma$---through a study of these quasiparticles, certainly not in the model-independent way we aspired to. Naively reinterpreting $\tau_\gamma$ as an effective coupling of the daughter particles to the photons---and further ignoring the vacuum mass $m$ in the dense plasma---would lead back to the usual bounds on millicharged particles~\cite{Davidson:2000hf}.

In conclusion, a massive photon sounds crazy and exotic, but it really is not. A massless photon is neither a theoretical prediction nor a necessity, but rather a phenomenological curiosity.
We should try to understand why this parameter in the Lagrangian (that we can just write down) is so small. This is similar to the strong $CP$ problem~\cite{Kim:2008hd}, and in both cases experiments so far have only come up with upper bounds for these parameters. Independent of its actual value, a nonzero photon mass immediately opens up the possibility of photon decay---even in the SM---which can, and should, also be constrained. Using the long-lived low-energy photons of the cosmic microwave background, we were able to derive the first direct bound on the photon lifetime in this Letter. Using the largest allowed value for the photon mass from other experiments, we find a lower limit of about $\unit[3]{yr}$ on the photon rest-frame lifetime. For photons in the visible spectrum, this corresponds to a lifetime around $\unit[10^{18}]{yrs}$.

A proper study of the challenging, but important, effects of the primordial plasma on this limit lies outside the scope of this Letter and will be left for future work.

\begin{acknowledgments}
The author thanks Werner Rodejohann and Joerg Jaeckel for discussions and comments.
This work was supported by the IMPRS-PTFS and the Max Planck Society through the Strategic Innovation Fund in the project MANITOP.
\end{acknowledgments}

\end{document}